\documentstyle[prl,aps,epsf]{revtex}
\input{epsf}

\begin{document}
\draft
\twocolumn[\hsize\textwidth\columnwidth\hsize\csname @twocolumnfalse\endcsname

\title{Static and dynamic coupling transitions 
of vortex lattices in disordered anisotropic superconductors}
\author{C. J. Olson$^{1}$, G.T. Zim{\' a}nyi$^{1}$,
A. Kolton$^{2}$ and N. Gr{\o}nbech-Jensen$^{3,4}$ 
} 
\address{$1. \ $Department of Physics, University of California, 
Davis, California 95616 \\ 
$2. \ $Centro Atomico Bariloche, 8400 S. C. de Bariloche, Rio Negro, 
Argentina \\
$3. \ $Department of Applied Science, University of California, Davis, 
California 95616 \\
$4. \ $NERSC, Lawrence Berkeley National Laboratory, Berkeley, 
California 94720}

\date{\today}
\maketitle
\begin{abstract}
We use three-dimensional molecular dynamics simulations of magnetically
interacting pancake vortices to study 
vortex matter
in disordered, highly anisotropic materials such as BSCCO.  
We observe a sharp 3D-2D transition from vortex lines to
decoupled pancakes as a function of relative interlayer coupling strength, 
with an accompanying large increase in the critical current remniscent
of a second peak effect. 
We find that decoupled pancakes, when driven, simultaneously 
recouple and order into a crystalline-like state at high drives.
We 
construct a dynamic phase diagram 
and show that the dynamic recoupling transition is associated
with a double peak in $dV/dI$.
\end{abstract}
\vspace{-0.1in}
\vspace{-0.2in}
\vskip1pc]

In highly anisotropic superconductors such as BSCCO,
the vortex lattice is composed of individual pancake vortices 
\cite{clem1}.  
These pancakes, which interact both magnetically
and through Josephson coupling, align
under certain conditions 
into elastic lines resembling those
found in 
isotropic superconductors.  
Three-dimensional (3D) line-like behavior 
has been observed in transformer
geometry measurements \cite{lopez2}, 
muon-spin-rotation \cite{lee3} 
and neutron scattering \cite{cubitt4}.
Under different conditions, however, 
the pancake vortices in each layer move independently of
the other layers, and the system acts like a
stack of independent thin film superconductors.
Such
two-dimensional (2D) behavior has also been seen
in transformer experiments 
\cite{busch5}.
Thus in layered superconductors
two different effective
dimensionalities of the vortex pancake lattice may appear,
each with different characteristic properties.

Layered superconductors exhibit a striking second peak in
magnetization measurements
\cite{cubitt4,SecondPeak6,tamegai7,yang8,zeldov9,khaykovich10},
corresponding to an abrupt increase in the
critical current of the material as the applied field is increased.
This second peak is especially sharp in BSCCO, as shown in
local Hall probe measurements 
\cite{tamegai7,zeldov9,khaykovich10}
and recent Josephson plasma frequency measurements \cite{chikumoto11}.
The lattice appears ordered at 
fields below the transition and
disordered above.
There seems to be no widely accepted agreement 
on the mechanism behind this transition,
although numerous scenarios have been suggested, including
vortex entanglement \cite{ertas15},
dislocation proliferations\cite{giamarchi13},
dynamic effects \cite{Dynamic16}, 
or a 3D to 2D transition in the vortex pancake lattice
\cite{cubitt4,tamegai7,yang17,glazman18,dimension19}.
The effect of strong
disorder on a possible 3D-2D transition is unclear,
and also it is not known how the transport properties would be
affected by a 3D-2D transition.

In 2D systems with uncorrelated pinning,
the vortex lattice can be dynamically reordered by an applied
driving current, passing from a glassy state at zero drive, through
plastic flow \cite{Bhattacharya20},
to a reordered state at high current 
\cite{reorder21,moon21a,olson22,kolton23}.  
In layered systems, when the between plane interactions of pancakes 
is weak enough that the pancakes are decoupled, the pancakes on each plane
should reorder when a high enough driving current is applied.  This
reordering may also be accompanied by a dynamically driven recoupling 
transition \cite{reorder21}, but it is unclear where this transition 
could be located in relation to the 2D reordering transition, and how
it is affected by the externally applied magnetic field.

To address the issue of possible 3D-2D transitions in
disordered anisotropic materials, 
we have developed a 
simulation which allows for decoupling by incorporating
the correct magnetic interactions between vortex pancakes.  
We show that a sharp 3D-2D transition occurs when
the relative strength of interlayer and intralayer pancake
interactions is varied, and that this transition  
is associated with a {\it sharp increase} in the critical 
current.
Furthermore, the system exhibits a rich array of dynamic 3D phases when
driven by a current \cite{reorder21}. 
We show that 2D decoupled pancakes can be dynamically recoupled in
a transition that occurs 
{\it simultaneously} with the dynamic ordering in each plane.
We construct a phase diagram as a function of interlayer coupling and
applied driving force, and show that a sharp,
experimentally observable second peak in $dV/dI$
is associated with the dynamic recoupling transition.

The magnetic interactions between pancakes in a layered material have
the same logarithmic form present in thin films, but
are highly anisotropic  
\cite{clem1,glazman18,pancakes24}. 
We have performed simulations in which pancakes in all layers
interact magnetically with long range interactions 
\cite{preprint24a,preprint24b}, 
in contrast to other simulations, which treated only
nearest layer interactions
\cite{wilkin25,vanO26}.
Our approach compliments calculations based on Lawrence-Doniach 
models \cite{ryu27a}.
The pancake interactions are long range both in and between planes, and are
treated according to Ref.\ \cite{ngj28}. 
Josephson coupling is neglected as a reasonable approximation
for materials in which the anisotropy $\gamma$ is sufficiently large
\cite{glazman18}.

\begin{figure}
\centerline{
\epsfxsize=3.5in
\epsfbox{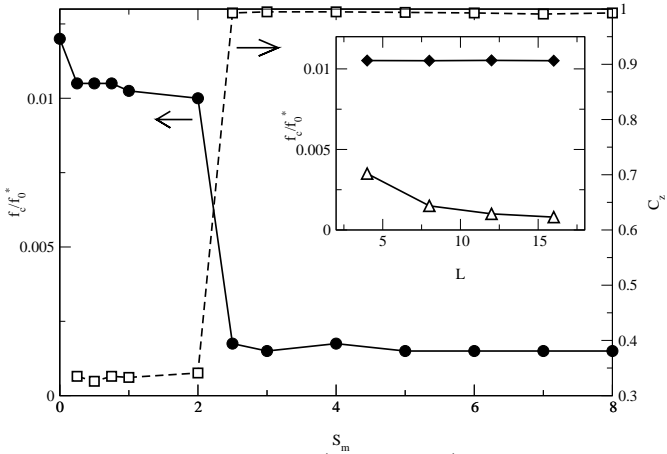}}
\caption{Critical current $f_c$ (filled circles) and
interlayer correlation $C_z$ (open squares) versus interlayer coupling 
strength $s_m$.  A transition from 2D to 3D behavior occurs between 
$s_m=2$
and $s_m=2.5$. 
Inset: Critical current $f_c$ for samples with $L=$ 4, 8, 12, and 16 layers.
Filled diamonds: $f_c$ for $s_m=0.5$, in the 2D regime.
Open triangles: $f_c$ for $s_m=5.0$, in the 3D regime.
}
\label{fig:jc}
\end{figure}

\hspace{-13pt}

The overdamped equation of motion, at $T=0$, for vortex $i$ is given by
$ {\bf f}_{i} = -\sum_{j=1}^{N_{v}}\nabla_{i} {\bf U}(\rho_{i,j},z_{i,j})
+ {\bf f}_{i}^{vp} + {\bf f}_{d}= {\bf v}_{i}$,
where $N_v$ is the number of vortices, and $\rho_{i,j}$ and 
$z_{i,j}$ are the distance
between pancakes $i$ and $j$ in cylindrical coordinates.
The system has periodic boundaries in-plane and open boundaries
in the $z$ direction.
The magnetic energy between pancakes is \cite{koshelev18a}
\begin{eqnarray}
{\bf U}(\rho_{i,j},0)=2d\epsilon_{0} 
\left((1-\frac{d}{2\lambda})\ln{\frac{R}{\rho}}
+\frac{d}{2\lambda} 
E_{1}\right) 
\nonumber
\end{eqnarray}
\begin{eqnarray}
{\bf U}(\rho_{i,j},z)=-s_{m}\frac{d^{2}\epsilon_{0}}{\lambda}
\left(\exp(-z/\lambda)\ln\frac{R}{\rho}- 
E_{2}\right) \nonumber
\end{eqnarray}
where
$E_{1} = 
\int^{\infty}_{\rho} d\rho^{\prime} \exp(\rho^{\prime}/\lambda)/\rho^{\prime}$,
$E_{2} = 
\int^{\infty}_{\rho} d\rho^{\prime} \exp(R/\lambda)/\rho^{\prime}$,
$R = 22.6 \lambda$, the maximum radial distance,
$\epsilon_{0} = \Phi_{0}^{2}/(4\pi\lambda)^{2}$, 
$\lambda$ is the London penetration depth,
$d=0.005\lambda$ is the interlayer spacing,
and $\xi$ is the coherence length.
We vary the relative strength of the interlayer coupling using the
prefactor $s_m$.
The uncorrelated pins are modeled by parabolic traps that are
randomly distributed in each layer.  The vortex-pin interaction is given by
${\bf f}_{i}^{vp} = \sum_{k=1}^{N_{p,L}} (f_{p}/\xi_{p})
({\bf r}_{i} - {\bf r}_{k}^{(p)}) \Theta (
(\xi_{p} - |{\bf r}_{i} - {\bf r}_{k}^{(p)} |)/\lambda)$,
where the pin radius $\xi_{p}=0.2\lambda$, the
pinning force $f_{p}=0.02f_{0}^{*}$, and $f_{0}^{*}=\epsilon_{0}/\lambda$.
The case of stronger pins is considered in \cite{preprint24a}.
The driving current must be increased slowly enough for the system to
equilibrate at each drive \cite{transverse29}.  Here $f_d$ is increased
by $0.00025f_{0}^{*}$ every 35000 time steps.
We have simulated a $16\lambda \times 16\lambda$ system 
with a vortex density of $n_v = 0.35/\lambda^2$ and a pin
density of $n_p = 1.0/\lambda^2$ in each of $L=8$ layers.  This
corresponds 
to $N_{v}=89$ vortices and $N_{p}=256$ pins per layer, with
a 

\begin{figure}
\centerline{
\epsfxsize=3.5in
\epsfbox{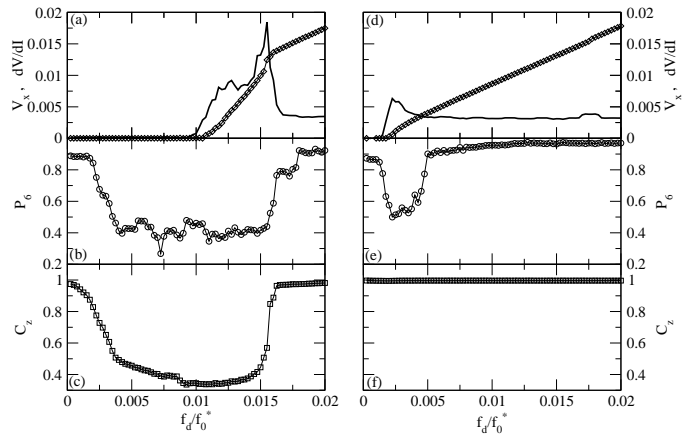}}
\caption{
(a) Diamonds: $V_x = (1/N_v)\sum_1^{N_v} v_x$; 
heavy line: $dV/dI$, for a sample with
$s_m=1.5$ that undergoes dynamic recoupling.  (b) $P_6$, the fraction of
sixfold coordinated vortices, for $s_m=1.5$.  (c) $C_z$, the
interlayer correlation, for $s_m=1.5$.  
(d) Diamonds: $V_x$; heavy line: $dV/dI$, for a sample with
$s_m=4.0$, which remains coupled at all drives.  (e) $P_6$, the fraction of
sixfold coordinated vortices, for $s_m=4.0$.  (f) $C_z$, the
interlayer correlation, for $s_m=4.0$.  
}
\label{fig:transition}
\end{figure}

\hspace{-13pt}
total of 712 pancake vortices.  We have checked for finite size 
effects on systems containing up to 42 layers and 3738 pancakes.

In the equilibrium state of the vortex lattice at zero drive, 
we find a {\it sharp 3D-2D transition} from vortex pancakes to
vortex lines when the strength $s_m$ of the interlayer coupling is decreased,
as shown in Fig.~\ref{fig:jc}.
We quantify the transition by measuring the spatial correlation
of pancakes in neighboring planes,
$C_z = 1 - \langle(|({\bf r}_{i,L}-{\bf r}_{i,L+1})|/(a_0/2))
\Theta( a_0/2 - |({\bf r}_{i,L}-{\bf r}_{i,L+1})|)\rangle$,
where 
$a_0$ is the vortex lattice constant.
A clear sharp drop in $C_z$ with decreasing $s_m$ 
appears in Fig.~\ref{fig:jc} at $s_m=2.0$
for a sample with $L=8$ layers.
The 3D-2D transition is accompanied by a large {\it increase} in the 
critical current $f_c$, as seen in Fig.~\ref{fig:jc}.  For weak interlayer
coupling, $s_m\le 2.0$, the different layers of the sample depin independently
and $f_c$ is close to the value it would have in a 2D sample. 
Here, $f_c$
is insensitive to the number of layers in the system, as can be seen
by comparing the data from samples with $L=$4 to 16 in the
inset of Fig.~\ref{fig:jc}.
In contrast, coupled lines of pancakes at $s_m>2.0$
average the
random pinning over their length and become 
much less effectively pinned.
Therefore $f_c$ decreases
with increasing number of layers as seen in the inset of Fig.~\ref{fig:jc}.  

When the magnetic field $H$ increases, pancakes within a plane are
brought closer together, but the distance between planes is unchanged.
Thus increasing $H$ corresponds to weakening the coupling between
planes by decreasing $s_m$.
Therefore, our results support the suggestion that the
sharp second peak observed in magnetization measurements results 
from a dimensional change in the vortex lattice from weakly pinned
3D vortex lines to strongly pinned 2D
pancakes as the magnetic field is 

\begin{figure}
\centerline{
\epsfxsize=3.5in
\epsfbox{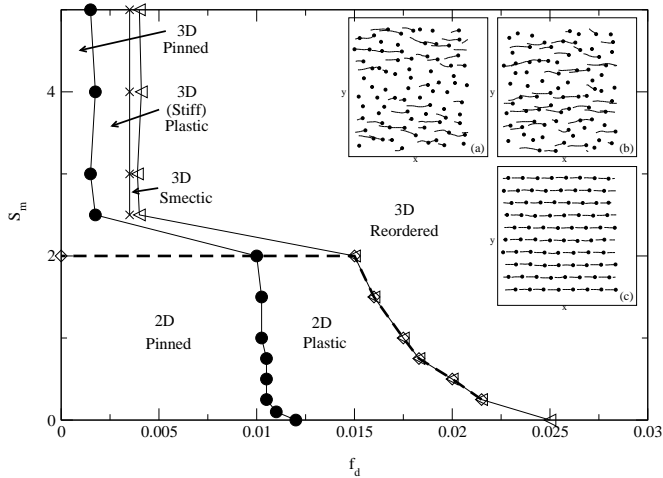}}
\caption{ Phase diagram for varying interlayer coupling $s_m$ and driving
force $f_d$.  Filled circles: depinning line.  X's: smectic transition
line (in 3D phase only). Triangles: in-plane reordering line.
Diamonds and heavy dashed line: recoupling line. 
For samples with $s_m\le 2.0$,
a dynamic transition into the 3D reordered state occurs at high drives.
Inset: Vortices (filled circles) and vortex trajectories (lines) 
at two different currents.  
(a) and (b): Vortex trajectories in top and bottom layers, 
respectively,
for a system with $s_m=1.5$ at $f_d/f_{0}^{*}=0.0125$, in the 2D plastic
flow regime.  The vortex positions and flow are different in each layer.
(c): Vortex trajectories
for a system with $s_m=1.5$ at $f_d/f_{0}^{*}=0.02$, in the 3D reordered
flow regime.}
\label{fig:phase}
\end{figure}

\hspace{-13pt}
increased \cite{blatter30}.  The 
behavior in $f_c$ that we observe indicates that {\it a sharp change in
transport properties can occur in a disordered system as a result of
a change in the effective dimensionality.}
Furthermore, the sharpness of the transition found both here and
in experiments suggests that the 3D-2D transition is 
{\it first order}.

We next consider the question of a possible dynamic recoupling transition
by applying a driving force.
When the vortex lattice begins to move for $f_d > f_c$,  it undergoes
plastic tearing due to the strong pinning in our sample.
For decoupled samples with $s_m\le 2.0$
each plane performs {\it independent 2D plastic flow}, as seen by the different
vortex positions and trajectories in the top and bottom layer of 
the sample shown in 
Fig.~\ref{fig:phase}(a,b).
When high drives are applied, however, the pancakes
{\it recouple} into lines and all planes begin to move in unison
[Fig.~\ref{fig:phase}(c)]. 

The reordering transition is shown in more detail in 
Fig.~\ref{fig:transition}(a-c).  Here, for a sample with $s_m=1.5$,
the recoupling transition in $C_z$ [Fig.~\ref{fig:transition}(c)] is 
sharp and occurs 
{\it simultaneously} with the in-plane reordering transition
indicated by $P_6$ [Fig.~\ref{fig:transition}(b)].
Furthermore, a sharp peak in $dV/dI$ appears at the transition, which
will be discussed in more detail below.
In contrast, as seen in
Fig.~\ref{fig:transition}(d-f), a sample with $s_m=4.0$ that is above the
static 2D-3D transition contains vortices that move as stiff
3D lines, and shows the same reordering
transitions seen in previous work on effectively two-dimensional systems 
\cite{olson22}.

We summarize the behavior of the system in the phase

\begin{figure}
\centerline{
\epsfxsize=3.5in
\epsfbox{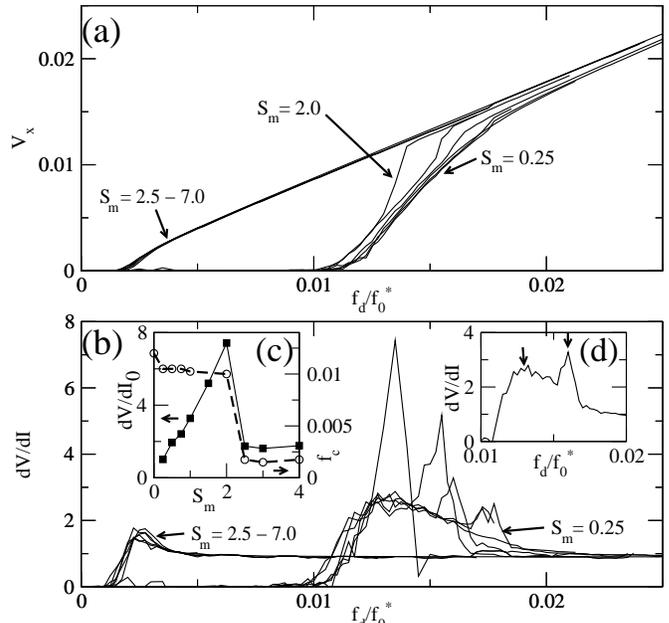}}
\caption{(a) $V_x(I)$ curves for samples with different values of $s_m$.
From right to left, $s_m=$0.25, 0.5, 0.75, 1.0, 1.5, 2.0, 2.5, 3.0, 
4.0, 5.0, 6.0, and 7.0.  Samples with $s_m > 2.0$ are coupled at all
drives; samples with $s_m \le 2.0$ undergo dynamic recoupling.
(b) $dV/dI$ curves for the same samples, from right to left, as listed
above.  
Samples with $s_m > 2.0$ show a single
peak in $dV/dI$ associated with the plastic flow
regime at currents just above the 3D depinning current of 
$f_d/f_{0}^{*} = 0.0015$.  Samples with $s_m \le 2.0$ show {\it two} peaks
in $dV/dI$.  
(c) Filled squares: The value of $dV/dI_0$, the maximum in the $dV/dI$ 
curve, for samples with different $s_m$.  $dV/dI_0$ increases significantly
as the transition value of $s_m$ is approached from below.  Open circles
and dashed line: $f_c$ versus $s_m$.
(d) $dV/dI$ for a single sample with $s_m=1.0$ illustrating the double
peak feature.
The first broad peak (left arrow) appears just above the 2D depinning
current of $f_d/f_{0}^{*}=0.010$ and is associated with 2D decoupled
plastic flow.  The second sharp peak (right arrow) 
is associated with the dynamic
recoupling transition.
}
\label{fig:vi}
\end{figure}

\hspace{-13pt}
diagram
of Fig.~\ref{fig:phase}.  The coupled vortices with $s_m>2.0$ undergo
plastic flow of stiff lines, pass through a smectic state, 
and finally reorder into a crystalline-like state at high drives.
The plastic flow and smectic regions shrink as the number of
layers is increased, and finally disappear, so that
for $L=16$ and $s_m=5$ we observe elastic
depinning directly into the 3D ordered state, with no plastic flow.
Decoupled vortices with $s_m \le 2.0$ depin into a 2D plastic flow phase
in which each layer moves independently.  
The vortices switch abruptly from 2D to 3D behavior at the recoupling
transition line, so they directly enter
the 3D ordered state.
The dynamic recoupling transition line and in-plane reordering
transition line fall on top of each other in the phase
diagram.  

As shown in Fig.~\ref{fig:vi}(b),
the dynamic recoupling and simultaneous reordering
are associated with a sharp {\it peak} in the $dV/dI$ curve.
This peak is {\it distinct} from the broader
peak in $dV/dI$ associated with plastic flow of the vortex lattice,
as indicated in Fig.~\ref{fig:vi}(d).  The sharp peak disappears into the
background value of $dV/dI$ and is not observed
when the interlayer coupling becomes too weak.
The height
of the reordering peak increases rapidly as the static 2D-3D transition 
value of $s_m$ is approached from below, as indicated in Fig.~\ref{fig:vi}(c),
and simultaneously the 
current at which the reordering peak 
appears shifts downwards towards the location of the broad plastic peak.
To understand the reordering peak, note that 
the 2D decoupled lattice depins 
at the high $f_c$ associated
with the 2D limit.  When the lattice recouples, it 
must cross from the low $V_x$ response in the
2D plastic limit up to the higher 3D elastic $V_x$ response curve at the
recoupling transition.  The location of the 
depinning transition is not affected by the value 
of $s_m$ since it is purely a 2D effect; however, the recoupling transition
is shifted to lower driving currents as $s_m$ increases.  
As a result, the rising V(I) curve
becomes steeper as the static transition point is
approached from below, as can be seen for $s_m=$1.0, 1.5, and 2.0 in
Fig.~\ref{fig:vi}(a).  The maximum value of $dV/dI$, which we call
$dV/dI_0$, correspondingly increases, as shown
in Fig.~\ref{fig:vi}(c).
When the static 2D-3D transition is crossed at $s_m=2.0$,
the second sharp peak disappears.
The behavior of this sharp peak in $dV/dI$, associated
with the dynamic recoupling transition, should be
experimentally observable in transport measurements performed at fields
approaching the second peak from above, when the vortex pancakes
are expected to be decoupled.

In summary, we have used a 3D molecular dynamics
simulation employing the magnetic interactions of pancake vortices
to study the dynamic phases of vortex matter
in disordered highly anisotropic materials such 
as BSCCO.  
As a function of the relative interlayer coupling strength,
we observe a sharp 3D-2D transition from vortex lines to
decoupled pancakes.
We find an {\it abrupt large increase in the critical current} as the
3D-2D line is crossed in a direction corresponding to increasing $H$, 
with decoupled pancakes being much more
strongly pinned.  
At driving currents well above
depinning, we find that the decoupled pancakes {\it simultaneously 
recouple and reorder} into a crystalline-like state at high drives.
We construct a phase diagram as a function of interlayer
coupling and 
show that the recoupling transition coincides with the
single-layer recrystallization transition.
We show that the recrystallization is associated with an 
experimentally observable double peak
in $dV/dI$ and that the peak height grows rapidly as the static
recoupling transition point is approached from below.

We acknowledge helpful discussions with L. N. Bulaevskii, 
D. Dominguez, C. Reichhardt, and R.T. Scalettar.
This work was supported by CLC and CULAR (LANL/UC),
by the NSF DMR 9985978, and by
the Director, Office of Adv.\ Scientific
Comp.\ Res., Div.\ of Math., Information, and 
Comp.\ Sciences, U.S.\ DoE contract DE-AC03-76SF00098.

\end{document}